\documentstyle[aps,eqsecnum,preprint,floats,epsf,epsfig]{revtex}
\begin{document}
\def\be{\begin{eqnarray}}
\def\en{\end{eqnarray}}
\def\non{\nonumber}
\def\la{\langle}
\def\ra{\rangle}
\def\a{{\cal A}}
\def\B{{\cal B}}
\def\c{{\cal C}}
\def\d{{\cal D}}
\def\e{{\cal E}}
\def\p{{\cal P}}
\def\t{{\cal T}}
\def\nc{N_c^{\rm eff}}
\def\vp{\varepsilon}
\def\drho{\bar\rho}
\def\deta{\bar\eta}
\def\vma{{_{V-A}}}
\def\vpa{{_{V+A}}}
\def\J{{J/\psi}}
\def\ov{\overline}
\def\Lqcd{{\Lambda_{\rm QCD}}}
\def\pr{{\sl Phys. Rev.}~}
\def\prl{{\sl Phys. Rev. Lett.}~}
\def\pl{{\sl Phys. Lett.}~}
\def\np{{\sl Nucl. Phys.}~}
\def\zp{{\sl Z. Phys.}~}
\def\lsim{ {\ \lower-1.2pt\vbox{\hbox{\rlap{$<$}\lower5pt\vbox{\hbox{$\sim$}
}}}\ } }
\def\gsim{ {\ \lower-1.2pt\vbox{\hbox{\rlap{$>$}\lower5pt\vbox{\hbox{$\sim$}
}}}\ } }

\font\el=cmbx10 scaled \magstep2{\obeylines \hfill July, 2002}

\vskip 1.5 cm

\centerline{\large\bf Nonresonant Three-body Decays of $D$ and $B$
Mesons}
\bigskip
\centerline{\bf Hai-Yang Cheng$^{1,2}$ and Kwei-Chou Yang$^{3}$}
\medskip
\centerline{$^1$ Institute of Physics, Academia Sinica}
\centerline{Taipei, Taiwan 115, Republic of China}
\medskip
\centerline{$^2$ C.N. Yang Institute for Theoretical Physics,
State University of New York} \centerline{Stony Brook, New York
11794}
\medskip
\centerline{$^3$ Department of Physics, Chung Yuan Christian
University} \centerline{Chung-Li, Taiwan 320, Republic of China}
\bigskip
\bigskip
\bigskip
\centerline{\bf Abstract}
\bigskip
{\small  Nonresonant three-body decays of $D$ and $B$ mesons are
studied. It is pointed out that if heavy meson chiral perturbation
theory (HMChPT) is applied to the heavy-light strong and weak
vertices and assumed to be valid over the whole kinematic region,
then the predicted decay rates for nonresonant charmless 3-body
$B$ decays will be too large and especially $B^-\to \pi^-K^+K^-$
greatly exceeds the current experimental limit. This can be
understood as chiral symmetry has been applied there twice beyond
its region of validity. If HMChPT is applied only to the strong
vertex and the weak transition is accounted for by the form
factors, the dominant $B^*$ pole contribution to the
tree-dominated direct three-body $B$ decays will become small and
the branching ratio will be of order $10^{-6}$. The decay modes
$B^-\to (K^-h^+h^-)_{\rm NR}$ and $\ov B^0\to(\ov K^0 h^+h^-)_{\rm
NR}$ for $h=\pi,K$ are penguin dominated. We apply HMChPT in two
different cases to study the direct 3-body $D$ decays and compare
the results with experiment. The preliminary FOCUS measurement of
the direct decay $D_s^+\to(\pi^+\pi^+\pi^-)_{\rm NR}$  may provide
the first indication of the importance of final-state interactions
for the weak annihilation process in nonresonant $D$ decays.
Theoretical uncertainties are discussed.

} \pagebreak

\section{Introduction}
The three-body decays of heavy mesons are in general dominated by
intermediate (vector or scalar) resonances, namely, they proceed
via quasi-two-body decays containing a resonance state and a
pseudoscalar meson. The analysis of these decays using the Dalitz
plot technique enables one to study the properties of various
resonances. The nonresonant contribution is usually a small
fraction of the total 3-body decay rate. Nevertheless, its study
is important for several reasons. First, the interference between
resonant and nonresonant decay amplitudes in $B$ decays may
provide information on the CP-violating phase angles
\cite{Deshpande,Fajfer1,Fajfer2,Deandrea1,Deandrea,Gardner}. For
example, the interference between $B^-\to(\pi^+\pi^-\pi^-)_{\rm
NR}$ and $B^-\to \chi_{c0}\pi^-$ could lead to a measurable $CP$
asymmetry characterized by the phase angle $\gamma$
\cite{Deshpande}, while the Dalitz plot analysis of
$B\to\rho\pi\to\pi\pi\pi$ allows one to measure the angle
$\alpha$. Second, an inadequate extraction of the nonresonant
contribution could yield incorrect measurements for the resonant
channels \cite{Bediaga}. Third, some of nonresonant 3-body $D$
decays have been measured. It is thus important to understand
their underlying mechanisms. Experimentally, it is hard to measure
the direct 3-body decays as the interference between nonresonant
and quasi-two-body amplitudes makes it difficult to disentangle
these two distinct contributions and extract the nonresonant one.

The direct three-body decays of mesons in general receive two
distinct contributions: one from the point-like weak transition
and the other from the pole diagrams which involve four-point
strong vertices. For $D$ decays, attempts of applying the
effective $SU(4)\times SU(4)$ chiral Lagrangian to describe the
$DP\to DP$ and $PP\to PP$ scattering at energies $\sim m_D$ have
been made by several authors \cite{Singer,KP,Cheng86,CC90,Botella}
to calculate the nonresonant $D$ decays, though in principle it is
not justify to employ the SU(4) chiral symmetry. As shown in
\cite{CC90,Botella}, the predictions of the nonresonant decay
rates in chiral perturbation theory are in general too small when
compared with experiment.

With the advent of heavy quark symmetry and its combination with
chiral symmetry \cite{Yan,Wise,Burdman}, the nonresonant $D$
decays can be studied reliably at least in the kinematical region
where the final pseuodscalar mesons are soft. Some of the direct
3-body $D$ decays were studied based on this approach
\cite{Zhang,Ivanov}.

Nonresonant charmless three-body $B$ decays have been recently
studied extensively based on heavy meson chiral perturbation
theory (HMChPT). However, the predicted decay rates are
unexpectedly large. For example, the branching ratio of $B^-\to
(\pi^+\pi^-\pi^-)_{\rm NR}$ is predicted to be of order $10^{-5}$
in \cite{Deshpande} and \cite{Fajfer1}. Therefore, it has a decay
rate larger than the two-body counterpart $B\to \pi\pi$. However,
it is found in \cite{Deandrea} that the dominant $B^*$ pole
contribution to the nonresonant $B^-\to \pi^+\pi^-\pi^-$ accounts
for a branching ratio of order only $1\times 10^{-6}$. Recently,
Belle \cite{Belle} and BaBar \cite{BaBar} have measured several
charmless three-body $B$ decays without making any assumptions on
the intermediate resonance states \cite{Belle}. The predicted
branching ratio of order $3\times 10^{-5}$ in \cite{Fajfer1} for
$B^-\to (K^-K^+\pi^-)_{\rm NR}$ already exceeds the upper limit
$1.2\times 10^{-5}$ by Belle \cite{Belle} and $7\times 10^{-6}$ by
BaBar \cite{BaBar} for resonant and nonresonant contributions.
Likewise, the predicted $\B(B^-\to\pi^+\pi^-\pi^-)_{\rm NR}\approx
4\times 10^{-5}$ in \cite{Fajfer1} is too large compared to the
limit $1.5\times 10^{-5}$ set by BaBar. Therefore, it is important
to reexamine and clarify the existing calculations.

The issue has to do with the applicability of HMChPT. In order to
apply this approach, two of the final-state pseudoscalars have to
be soft. The momentum of the soft pseudoscalar should be smaller
than the chiral symmetry breaking scale $\Lambda_\chi\sim 830$
MeV. For 3-body charmless $B$ decays, the available phase space
where chiral perturbation theory is applicable is only a small
fraction of the whole Dalitz plot. Therefore, it is not justified
to apply chiral and heavy quark symmetries to a certain kinematic
region and then generalize it to the region beyond its validity.
In order to have a reliable prediction for the {\it total rate} of
direct 3-body decays, one should try to utilize chiral symmetry to
a minimum. Therefore, we will apply HMChPT only to the strong
vertex and use the form factors to describe the weak vertex. In
contrast, for direct 3-body $D$ decays, the allowed phase space
region where HMChPT is applicable can be a dominant one for some
decay modes.

The paper is organized as follows. After introducing the effective
Hamiltonian in Sec. II we proceed to discuss the difficulties with
HMChPT when applying it to describe the 3-body nonresonant $B$
decays in the whole Dalitz plot and its possible remedy. The full
amplitude for the penguin-dominated $B^-\to K^-\pi^+\pi^-$ is
worked out as an example. The direct 3-body $D$ decays are
discussed in Sec. III. Discussions of theoretical uncertainties
and conclusions are presented in Sec. IV.

\section{Nonresonant three-body decays of $B$ mesons}
\subsection{Hamiltonian}
The relevant effective $\Delta B=1$ weak Hamiltonian for charmless
hadronic $B$ decays is \be {\cal H}_{\rm eff}(\Delta B=1) &=&
{G_F\over\sqrt{2}}\Bigg\{ V_{ub}V_{uq}^*
\Big[c_1(\mu)O_1^u(\mu)+c_2(\mu)O_2^u(\mu)\Big]+V_{cb}V_{cq}^*\Big[c_1(\mu)
O_1^c(\mu)+c_2(\mu)O_2^c(\mu)\Big]  \non \\
&& -V_{tb}V_{tq}^*\sum^{10}_{i=3}c_i(\mu)O_i(\mu)\Bigg\}+{\rm
h.c.}, \en where $q=d,s$, and \be && O_1^u= (\bar ub)_\vma(\bar
qu)_\vma, \qquad\qquad\qquad\qquad~~
O_2^u = (\bar u_\alpha b_\beta)_\vma(\bar q_\beta u_\alpha)_\vma, \non \\
&& O_1^c= (\bar cb)_\vma(\bar qc)_\vma,
\qquad\qquad\qquad\qquad~~~
O_2^c = (\bar c_\alpha b_\beta)_\vma(\bar q_\beta c_\alpha)_\vma, \non \\
&& O_{3(5)}=(\bar qb)_\vma\sum_{q'}(\bar q'q')_{\vma(\vpa)},
\qquad  \qquad O_{4(6)}=(\bar q_\alpha b_\beta)_\vma\sum_{q'}(\bar
q'_\beta q'_\alpha)_{
\vma(\vpa)},   \\
&& O_{7(9)}={3\over 2}(\bar qb)_\vma\sum_{q'}e_{q'}(\bar
q'q')_{\vpa(\vma)},
  \qquad O_{8(10)}={3\over 2}(\bar q_\alpha b_\beta)_\vma\sum_{q'}e_{q'}(\bar
q'_\beta q'_\alpha)_{\vpa(\vma)},   \non \en with $O_3$--$O_6$
being the QCD penguin operators, $O_{7}$--$O_{10}$ the electroweak
penguin operators and $(\bar q_1q_2)_{_{V\pm A}}\equiv\bar
q_1\gamma_\mu(1\pm \gamma_5)q_2$. The scale dependent Wilson
coefficients calculated at next-to-leading order are
renormalization scheme dependent.   In the factorization approach
the decay amplitude has the form
 \be
 A(B\to M_1M_2M_3)\propto \sum a_i\la M_1M_2M_3|O_i|B\ra,
 \en
where the coefficients $a_i$ are renormalization scale and
$\gamma_5$-scheme independent. In ensuing calculations we will
employ the values of $a_i$ listed in \cite{Du}. For $D$ decays we
will use
  \be \label{Da12}
 a_1=1.20\,, \qquad\qquad a_2=-0.67\,.
 \en

\subsection{Difficulties with heavy meson chiral perturbation theory for nonresonant $B$ decays}
The nonresonant three-body $B$ decays have been studied in two
distinct methods, though both are based on heavy quark symmetry.
One relies heavily on chiral perturbation theory to evaluate the
3-body matrix elements \cite{Fajfer1,Fajfer2,Fajfer3}, whereas the
use of chiral symmetry is restricted to the strong vertex for the
other case \cite{Deshpande,Deandrea}. The resulting decay rates
can be different by one to two orders of magnitude.

Let us first recapitulate the approach of heavy meson chiral
perturbation theory \cite{Yan,Wise,Burdman} and consider the decay
mode $B^-\to (K^-K^+\pi^-)_{\rm NR}$ as an illustration. Since
this decay is tree dominated, we will focus on the dominant
contribution from the four-quark operator $O_1$
 \be
 A(B^-\to K^-(p_1)K^+(p_2)\pi^-(p_3)) = {G_F\over\sqrt{2}}
 V_{ub}V_{ud}^*\,a_1\la K^-K^+\pi^-|O_1|B^-\ra.
 \en
Under the factorization approximation,
 \be
 \la K^-K^+\pi^-|O_1|B^-\ra &=& \la \pi^-|(\bar du)_\vma|0\ra\la
 K^-K^+|(\bar u b)_\vma|B^-\ra \non \\ &+& \la K^-K^+\pi^-|(\bar du)_\vma|0\ra
 \la 0|(\bar u b)_\vma|B^-\ra.
 \en
The second term on the right hand side corresponds to weak
annihilation and it is expected to be helicity suppressed. As we
shall see below, it indeed vanishes in the chiral limit.

The three-body matrix element $\la K^-K^+|(\bar u b)_\vma|B^-\ra$
has the general expression
 \cite{Lee}
 \be
 \la K^-(p_1)K^+(p_2)|(\bar u b)_\vma|B^-(p_B)\ra &=&
 ir(p_B-p_1-p_2)_\mu+i\omega_+(p_2+p_1)_\mu \non \\ &+& i\omega_-(p_2-p_1)_\mu
 +h\epsilon_{\mu\nu\alpha\beta}p_B^\nu (p_2+p_1)^\alpha
 (p_2-p_1)^\beta,
 \en
where $r$, $\omega_\pm$ and $h$ are the unknown form factors. When
pseudoscalar mesons are soft, the heavy-to-light current in the
heavy quark limit can be expressed in terms of a heavy meson and
light pseudoscalar mesons \cite{Wise,Yan}. The weak current
$L^\mu_a=\bar q_a\gamma_\mu(1-\gamma_5)Q$, when written in terms
of a heavy meson and light pseudoscalars, has the form \cite{Wise}
 \be \label{Qqcurrent}
 L^\mu_a={if_{H_b}\sqrt{m_{H_b}}\over 2}\,{\rm Tr}
 [\gamma^\mu(1-\gamma_5)H_b\xi_{ba}^\dagger]
 \en
to the lowest order in the light meson derivatives, where $H_a$
contains the pseudoscalar meson $P_a$ and the vector-meson field
$P^*_{a\mu}$:
 \be
 H_a=\sqrt{m_{H_a}}\,{1+v\!\!\!/\over 2}(P_{a\mu}^*\gamma^\mu-P_a\gamma_5),
 \en
where $v$ is the velocity of the heavy meson and $\xi^2$ is equal
to the unitary matrix $U$ which describes the Goldstone bosons.
The general expression of the matrix $U$ up to the fourth order in
the meson matrix $\phi$ is \cite{Cheng88}
 \be
 U=1+2i{\phi\over f_\pi}-2{\phi^2\over f_\pi^2}-ia_3{\phi^3\over
 f_\pi^3}+2(a_3-1){\phi^4\over f_\pi^4}+\cdots,
 \en
where $a_3$ indicates the nonlinear chiral realization and it has
the well-known value ${4\over 3}$ in the usual exponential
expression for $U$, namely, $U={\rm exp}(i2\phi/f_\pi)$. Here we
do not specify the value of $a_3$ in order to demonstrate that the
physical quantity is independent of the choice of chiral
realization, i.e. the value of $a_3$. The traceless meson matrix
$\phi$ reads
 \be
 \phi=\left(\matrix{ {\pi^0\over\sqrt{2}}+{\eta\over\sqrt{6}} &
 \pi^+ & K^+ \cr  \pi^- &
 -{\pi^0\over\sqrt{2}}+{\eta\over\sqrt{6}}& K^0 \cr K^- & \ov K^0
 & -\sqrt{2\over 3}\,\eta \cr}\right).
 \en

\begin{figure}[t]
\vspace{-1cm} \hskip 3.5cm
  \psfig{figure=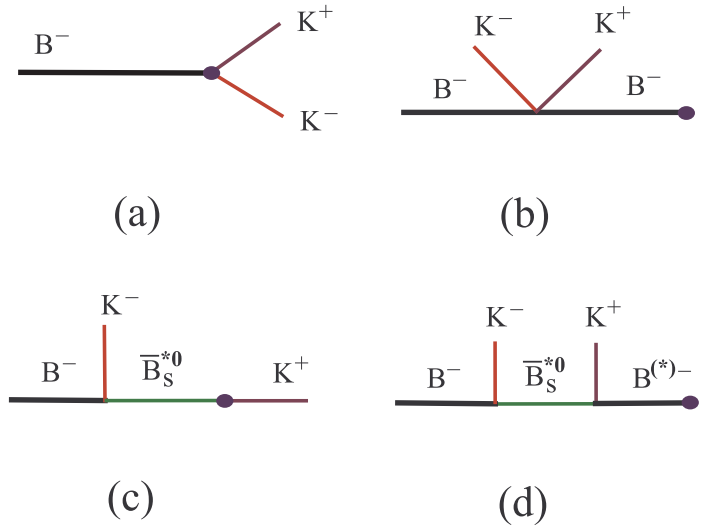,width=10cm}
\vspace{00.5cm}
    \caption[]{\small Point-like and pole diagrams responsible for the $B^-\to K^-K^+$
    matrix element of the current $\bar u\gamma_\mu(1-\gamma_5)b$,
    where the symbol $\bullet$ denotes an insertion of the current.}
\end{figure}

To compute the form factors $r$, $\omega_\pm$ and $h$, one needs
to consider not only the point-like contact diagram, Fig. 1(a),
but also various pole diagrams shown in Fig. 1. The heavy meson
chiral Lagrangian given in \cite{Yan,Wise,Burdman} is needed to
compute the strong $B^*BP$, $B^*B^*P$ and $BBPP$ vertices. The
results for the form factors are \cite{Lee,Fajfer1}
 \be \label{r&omega}
 \omega_+ &=& -{g\over f_\pi^2}\,{f_{B_s^*}m_{B_s^*}\sqrt{m_Bm_{B_s^*}}\over
 t-m_{B_s^*}^2}\left[1-{(p_B-p_1)\cdot p_1\over
 m_{B_s^*}^2}\right]+{f_B\over 2f_\pi^2}, \non \\
 \omega_- &=& {g\over f_\pi^2}\,{f_{B_s^*}m_{B_s^*}\sqrt{m_Bm_{B_s^*}}\over
 t-m_{B_s^*}^2}\left[1+{(p_B-p_1)\cdot p_1\over
 m_{B_s^*}^2}\right], \non \\
 r &=& {f_B\over 2f_\pi^2}-{f_B\over
 f_\pi^2}\,{p_B\cdot(p_2-p_1)\over
 (p_B-p_1-p_2)^2-m_B^2}+{2gf_{B_s^*}\over f_\pi^2}\sqrt{m_B\over
 m_{B_s^*}}\,{(p_B-p_1)\cdot p_1\over t-m_{B_s^*}^2} \non \\
 &-& {4g^2f_B\over f_\pi^2}\,{m_Bm_{B_s^*}\over
 (p_B-p_1-p_2)^2-m_B^2}\,{p_1\!\cdot\!p_2-p_1\!\cdot\!(p_B-p_1)\,p_2\!\cdot\!
 (p_B-p_1)/m_{B_s^*}^2 \over t-m_{B_s^*}^2 },
 \en
with $t\equiv (p_B-p_1)^2=(p_2+p_3)^2$. Note that the term
$f_B/(2f_\pi^2)$ comes from the point-like diagram, while the
other terms in $\omega_+$ and $\omega_-$ arise from the $B_s^*$
pole contributions in Fig. 1. The decay amplitude then reads
 \be \label{piKK1}
 &&A(B^-\to K^-(p_1)K^+(p_2)\pi^-(p_3))_{\rm NR}= -{G_F\over\sqrt{2}}
 V_{ud}V_{ub}^* a_1{f_\pi\over 2}  \\ && \qquad\qquad  \times
 \left\{2m_3^2r+(m_B^2-s-m_3^2)\omega_++(2t+s-m_B^2-2m_2^2-m_3^2)
 \omega_-\right\},  \non
 \en
with $s\equiv(p_B-p_3)^2=(p_1+p_2)^2$. It is clear that the
contribution due to the form factor $r$ is proportional to
$m_\pi^2$ and hence negligible. For the strong coupling  $g$,
which will be introduced again below, we shall employ the value of
$g=0.59\pm0.01\pm0.07$ as extracted from the recent CLEO
measurement of the $D^{*+}$ decay width \cite{CLEOg}.

The decay rate of $B^-\to K^-K^+\pi^-$ is then given by
 \be
 \Gamma(B^-\to K^-K^+\pi^-)=\,{1\over
 (2\pi)^3}\,{1\over 32m_B^3}\int_{t_{\rm min}}^{t_{\rm max}}
 \int_{s_{\rm min}}^{s_{\rm max}}|A|^2\,ds\,dt.
 \en
For a given $s$,  the upper and lower bounds of $t$ is fixed. If
(\ref{piKK1}) is applicable to the whole kinematical region, then
$s_{\rm min}=(m_1+m_2)^2$ and $s_{\rm max}=(m_B-m_3)^2$, and the
branching ratio of $B^-\to K^- K^+\pi^-$ is found to be
 \be
 \B(B^-\to K^-K^+\pi^-)_{\rm NR}=\cases{2.8\times 10^{-5} &
 from~the~contact~term~only,
  \cr  6.7\times 10^{-5} & from~the~$B^*$~pole~only,  \cr
  1.7\times 10^{-4} & total. \cr  }
  \en
This is already above the upper limit of $7.5\times 10^{-5}$ set
by CLEO \cite{CLEO}, and it greatly exceeds the experimental limit
$1.2\times 10^{-5}$ reported recently by Belle \cite{Belle} and
$7\times 10^{-6}$ by BaBar \cite{BaBar}, recalling that both Belle
and BaBar do not make any assumptions about intermediate
resonances. In other words, the upper bound on the nonresonant
$B^-\to \pi^- K^+K^-$ is presumably much less than $1\times
10^{-5}$ after subtracting resonant contributions. Therefore, it
is very likely that the branching ratio of direct $B\to PPP$
decays is overestimated by one to two orders of magnitude in this
approach.

The dominant contributions to the direct $B^-\to K^-K^+\pi^-$ come
from the $B^*$ pole and the point-like weak transition term
$f_B/f_\pi^2$. Since the chiral representation for the
heavy-to-light current is valid only for low momentum
pseudoscalars, the contact contribution from $\la \pi^-|(\bar
du)|0\ra\la K^+K^-|(\bar ub)|B^-\ra$ and the weak $B^*$ to $K$
transition in the $B^*$ pole diagrams are reliable only in the
kinematic region where $K^+$ and $K^-$ are soft. Therefore, the
available phase space where chiral perturbation theory is
applicable is very limited. It is claimed in
\cite{Fajfer1,Fajfer2,Fajfer3} that if the usual HQET Feynman
rules for the vertices near and outside the zero-recoil region but
the complete propagators instead of the usual HQET propagator are
used, then the model is applicable to the whole Dalitz plot.
However, as shown above, this will lead to too large decay rates
in disagreement with experiment. Therefore, in order to estimate
the nonresonant rates for the whole kinematic region, one should
try to apply chiral symmetry to a minimum or some assumptions have
to be made to extrapolate chiral symmetry results to the whole
phase space.

\subsection{$B^*$ pole contribution}
As discussed before, the direct contact contribution to the matrix
element $\la K^+K^-|(\bar ub)_\vma|B^-\ra$ as characterized by the
$f_B/f_\pi^2$ term is valid only in the chiral limit, and hence we
will not consider its contribution when computing the total decay
rate. As for the $B^*$ pole contribution, we shall try to avoid
the use of chiral symmetry when computing the $B^*_s$ to $K$ weak
transition; that is, we shall not use Eq. (\ref{Qqcurrent}) to
evaluate the matrix element of the $B^*\to P$ transition and we
apply HMChPT only to the strong vertex and use form factors to
describe the weak vertices. In this way, the soft meson limit is
applied only once rather than twice.

For the tree-dominated decay $B^-\to K^-K^+\pi^-$, the $B^*_s$
pole contribution is\footnote{The pole contribution from the
scalar meson $B_0$ and the effect of the decay width in the
propagator have been considered in \cite{Deandrea1}. We find these
effects are small.}
 \be
 A^\mu_{B^*_s\pi K}\,{i(-g_{\mu\nu}+p_{B^*_s\mu}p_{B^*_s\nu})/m_{B^*_s}^2\over
 p_{B^*_s}^2-m_{B^*_s}^2}\,A_{BB^*_sK}^\nu.
 \en
The general expression for $A_{BB^*_sK}^\nu$ is
 \be \label{BVBPcoup}
 \vp_\nu A_{BB^*_sK}^\nu=\la K^-(q)B^+(p_B)|B^{*0}_s(p_{B^*_s})\ra=g_{BB^*_s K}(\vp\cdot
 q).
 \en
In  heavy quark and chiral limits, the strong coupling $g_{BB^*_s
K}$ is determined to be \cite{Yan,Wise,Burdman}
 \be \label{g}
 g_{BB^*_s K}={2g\over f_\pi}\sqrt{m_B m_{B^*_s}},
 \en
where $g$ is a heavy-flavor independent strong coupling and its
sign is positive \cite{Yan}. It should be stressed that the
relation (\ref{g}) is valid only when the kaon is soft. Under the
factorization approximation
 \be
 \vp_\mu A^\mu_{B^*_s\pi
 K}={G_F\over\sqrt{2}}V_{ub}V_{ud}^*\,a_1\la\pi^-(p_3)|(\bar
 du)_\vma|0\ra\la K^+(p_2)|(\bar ub)_\vma|\ov B_s^{*0}\ra.
 \en
Heavy quark symmetry is then applied to relate the matrix element
of $\ov B_s^{*0}\to K^+$ to $\ov B^0_s\to K^+$ \cite{Deshpande}:
 \be \label{m.e.1}
 \la K^+(p_K)|(\bar ub)_\vma|\ov
 B_s^{*0}(p_{B_s^*})\ra &=& T_1i\epsilon_{\mu\nu\alpha\beta}\vp^\nu p_{B_s^*}^\alpha
 p_K^\beta-T_2m_{B_s^*}^2\vp_\mu -T_3(\vp\cdot p_K)(p_{B_s^*}+p_K)_\mu
 \non \\ &-& T_4(\vp\cdot p_K)(p_{B_s^*}-p_K)_\mu, \non \\
 \la K^+(p_K)|(\bar ub)_\vma|\ov
 B_s^{0}(p_{B_s})\ra &=& f_+(p_{B_s}+p_K)_\mu+f_-(p_{B_s}-p_K)_\mu,
 \en
with $\vp_\mu$ being the polarization vector of $\ov B^*_s$. The
result is (see e.g. \cite{Deshpande})\footnote{It is most
convenient to apply the interpolating field method for heavy
mesons (see e.g. \cite{Yan}), namely, $|\ov B^*\ra=\bar
h_v^{(b)}\vp\!\!\!/q$ and $|\ov B\ra=\bar h_v^{(b)}i\gamma_5 q$,
to relate the $B^*\to P$ form factors to those of $B\to P$. The
matrix element $\la\pi^+|(\bar ub)_\vma|\ov B^0\ra$ is also
evaluated in \cite{Deandrea1} using the relativistic potential
model. However, only the form factor $T_2$ is calculated there.}
 \be \label{Ti}
 && T_1= -{f_+-f_-\over m_B}, \qquad T_2={1\over
 m_B^2}\left[(f_++f_-)m_B+(f_+-f_-){p_{B^*}\cdot p_K\over m_B}\right], \non \\
 && T_3=-{f_+-f_-\over 2m_B}, \qquad\quad T_4=T_3.
 \en
In terms of the form factors $F_{1,0}^{B_sK}$ defined by
\cite{BSW}
 \be
  \la K^+(p_K)|(\bar ub)_\vma|\ov  B_s^{0}(p_{B})\ra=
  (p_B+p_K)_\mu F_1^{B_sK}(q^2)+{m_{B_s}^2-m_K^2\over
  q^2}q_\mu[F_0^{B_sK}(q^2)-F_1^{B_sK}(q^2)]
  \en
with $q_\mu=(p_B-p_K)_\mu$, we obtain
 \be \label{f12}
 f_+=F_1^{B_sK}, \qquad f_-=-{m_B^2\over
 m_\pi^2}\,F_1^{B_sK}\left(1-{F_0^{B_sK}\over F_1^{B_sK}}\right),
 \en
and
 \be \label{Anu}
 \vp_\mu A^\mu_{B^*_s\pi K} &=& -i{G_F\over\sqrt{2}}V_{ub}V_{ud}^*\,a_1f_\pi(\vp\cdot
 p_3)F_1^{B_s K}(m_\pi^2) \non \\
 &\times& \left[ m_B+{t\over m_B}-m_B{m_B^2-t\over
 m_\pi^2}\left(1-{F_0^{B_s K}(m_\pi^2)\over
 F_1^{B_sK}(m_\pi^2)}\right)\right].
 \en
Hence, the $B_s^*$ pole contribution to $B^-\to K^-K^+\pi^-$ is
 \be \label{piKKpole}
 A(B^-\to K^-(p_1)K^+(p_2)\pi^-(p_3))_{\rm pole} &=& {G_F\over\sqrt{2}}V_{ub}V_{ud}^*\,a_1F_1^{B_s
 K}(m_\pi^2){g\over t-m_{B_s^*}^2}\,\sqrt{m_Bm_{B_s^*}} \non \\
  &\times& \left[ m_B+{t\over m_B}-m_B{m_B^2-t\over
 m_\pi^2}\left(1-{F_0^{B_s K}(m_\pi^2)\over
 F_1^{B_sK}(m_\pi^2)}\right)\right]   \\
 &\times&
 \left[s+t-m_B^2-m_2^2+{(t-m_2^2+m_3^2)(m_B^2-t-m_1^2)\over
 2m_{B^*_s}^2}\right].  \non
 \en
Using the Melikov-Stech model \cite{MS} for the $B_s\to K$ form
factors, the branching ratio due to the $B_s^*$ pole is found to
be of order $1.8\times 10^{-6}$, which is consistent with the
upper limit $1.2\times 10^{-5}$ set by Belle \cite{Belle} and
$7\times 10^{-6}$ by BaBar \cite{BaBar}.

In contrast, the matrix element of $\ov B_s^{*0}\to K^+$ in HMChPT
has the form
 \be \label{m.e.2}
 \la K^+(p_K)|(\bar ub)_\vma|\ov B_s^{*0}(p_{B_s^*})\ra = {f_{B^*_s}\over
 f_\pi}m_{B^*_s}\vp_\mu.
 \en
Comparing this with Eqs. (\ref{m.e.1}) and (\ref{Ti}) it is clear
that in the heavy quark and chiral ($p_K\to 0$) limits, only the
form factor $T_2$ contributes with
 \be \label{T2a}
 m_BT_2=-{f_{B_s^*}\over f_\pi}={f_{B_s}\over f_\pi} \qquad{\rm
 in~heavy~quark~and~chiral~limits},
 \en
where use of Eq. (\ref{f12}) has been made. However, beyond the
chiral limit, all $T_2$, $T_3$ and $T_4$ contribute and
 \be \label{T2}
 m_BT_2=\,F_1^{B_sK}(m_\pi^2)\left[1+{t-m_\pi^2+m_K^2\over 2m_B^2}
 -{2m_B^2-t+m_\pi^2-m_K^2\over 2m_\pi^2}
 \left(1-{F_0^{B_s K}(m_\pi^2)\over
 F_1^{B_sK}(m_\pi^2)}\right)\right]
 \en
in the heavy quark limit. Since $F_1^{B_sK}(0)=0.31$ in the MS
form-factor model \cite{MS}, it is evident that the form factor
$T_2$ inferred from Eq. (\ref{T2}) is much smaller than that
implied by Eq. (\ref{T2a}), namely, $T_2=f_{B_s}/f_\pi=1.6$ for
$f_{B_s}=190$ MeV. This explains why the prediction based on
HMChPT is too large by one to two orders of magnitude compared to
the $B^*$ pole contribution which relies on chiral symmetry only
at the strong vertex.

The previous estimate of $B^-\to(\pi^+\pi^-\pi^-)_{\rm NR}$ by
Deshpande {\it et al.} \cite{Deshpande}  based on the $B^*$ pole
contribution gives a branching ratio of order $2\times 10^{-5}$
for $F_1^{B\pi}(0)=0.333$ and $g=0.60$ (case 1 in
\cite{Deshpande}). This is larger than our result $3.0\times
10^{-6}$ (see Table I below) by one order of magnitude. It can be
traced back to the square bracket term in Eq. (\ref{Anu}) for the
analogous $\vp_\nu A^\nu_{B^*\pi\pi}$ term where Deshpande {\it et
al.} obtained
 \be
 \left[{3\over 2} m_B+{t\over 2m_B}-{m_B\over 2}{m_B^2-t\over
 m_\pi^2}\left(1-{F_0^{B\pi}(m_\pi^2)\over
 F_1^{B\pi}(m_\pi^2)}\right)\right],
 \en
to be compared with
  \be
 \left[m_B+{t\over m_B}-m_B{m_B^2-t\over
 m_\pi^2}\left(1-{F_0^{B\pi}(m_\pi^2)\over
 F_1^{B\pi}(m_\pi^2)}\right)\right]
 \en
in our case. Numerically, the decay rate obtained by Deshpande
{\it et al.} is larger than ours by a factor of 3 when the same
$B\to\pi$ form factors are employed. Note that the $B^*$ pole
contribution to $B^-\to \pi^+\pi^-\pi^-$ is found to be $1.8\times
10^{-6}$ (for $g=0.6$) in \cite{Deandrea} and $2.7\times 10^{-6}$
in \cite{Gardner}. Therefore, our result is consistent with them.

\subsection{Full contributions}
In the previous subsections we have only considered the dominant
contribution to the tree-dominated $B$ decay from the operator
$O_1$. In the following we discuss the full amplitude for the
direct 3-body $B$ decay and choose the penguin-dominated decay
$B^-\to \pi^-\pi^+K^-$ as an example. The factorizable amplitude
reads
 \be \label{pipiK}
 A(B^-\to \pi^-(p_1)\pi^+(p_2)K^-(p_3)) &=& {G_F\over\sqrt{2}}\Bigg\{
 V_{ub}V_{us}^*\Big[a_1\la K^-|(\bar su)_\vma|0\ra\la
 \pi^+\pi^-|(\bar u b)_\vma|B^-\ra \non \\ &+& \la \pi^-\pi^+K^-|(\bar su)_\vma|0\ra
 \la 0|(\bar u b)_\vma|B^-\ra\Big] \non \\
 &+&  a_2\la \pi^-\pi^+|(\bar uu)_\vma|0\ra\la
 K^-|(\bar s b)_\vma|B^-\ra  \\
 &+& {3\over 2}(a_7+a_9)\la \pi^-\pi^+|(e_u\bar uu+e_d\bar dd)_\vma|0\ra\la
 K^-|(\bar s b)_\vma|B^-\ra  \non \\
 &-& V_{tb}V_{ts}^*\Big[a_4\la \pi^-\pi^+K^-|O_4|B^-\ra+
 a_6\la \pi^-\pi^+K^-|O_6|B^-\ra \non \\
 &+&  (4\to 10)+(6\to 8)\Big]\Bigg\}. \non
 \en
Under the factorization approximation, the matrix element of $O_4$
is
 \be
 \la \pi^-\pi^+K^-|O_4|B^-\ra &=& \la K^-|(\bar su)_\vma|0\ra\la
 \pi^-\pi^+|(\bar u b)_\vma|B^-\ra \non \\ &+& \la \pi^+K^-|(\bar
 sd)_\vma|0\ra\la \pi^-|(\bar d b)_\vma|B^-\ra \non \\
 &+& \la \pi^-\pi^+K^-|(\bar su)_\vma|0\ra\la 0|(\bar u
 b)_\vma|B^-\ra.
 \en
In Eq. (\ref{pipiK}) the two-body matrix element $\la
\pi^+K^-|(\bar
 sd)_\vma|0\ra$ has the form
 \be
 \la \pi^+(p_2)K^-(p_3)|(\bar sd)_\vma|0\ra &=& \la \pi^+(p_2)|(\bar
 sd)_\vma|K^+(-p_3)\ra = (p_3-p_2)_\mu F_1^{K\pi}(t) \non \\
 &+& {m_K^2-m_\pi^2\over
 t}(p_3+p_2)_\mu\left[-F_1^{K\pi}(t)+F_0^{K\pi}(t)\right],
 \en
where we have taken into account the sign flip arising from
interchanging the operators $s\leftrightarrow d$. The other
two-body matrix element $\la\pi^+\pi^-|(\bar uu)_\vma|0\ra$ can be
related to the pion matrix element of the electromagnetic current
 \be
 \la\pi^+(p)|J_\mu^{\rm em}|\pi^+(p')\ra &=&
 (p+p')_\mu F^{\pi\pi}(q^2),
 \non \\
  \la\pi^-(p)|J_\mu^{\rm em}|\pi^-(p')\ra &=&
  -(p+p')_\mu F^{\pi\pi}(q^2),
 \en
with $q^2=(p'-p)^2$ and $J_\mu^{\rm em}={2\over 3}\bar u\gamma_\mu
u-{1\over 3}\bar d\gamma_\mu d+\cdots$. The electromagnetic form
factor $F^{\pi\pi}$ is normalized to unity at $q^2=0$. Applying
the isospin relations yields
 \be
 \la \pi^+(p)|\bar u\gamma_\mu u|\pi^+(p')\ra=\la \pi^-(p)|\bar d\gamma_\mu d|\pi^-(p')\ra=
 (p+p')_\mu F^{\pi\pi}(q^2).
 \en

As for the three-body matrix element $\la \pi^-\pi^+K^-|(\bar
su)_\vma|0\ra$, one may argue that it vanishes in the chiral limit
owing to the helicity suppression. To see this is indeed the case,
we first assume that the kaon and pions are soft. The weak current
can be expressed in terms of the chiral representation derived
from the chiral Lagrangian
 \be \label{chiLang}
 {\cal L}=\,{f_\pi^2\over 8}{\rm Tr}(\partial_\mu U\partial^\mu
 U^\dagger)+{f_\pi^2\over 8}{\rm Tr}(MU^\dagger+U^\dagger M).
 \en
The weak current $J_\mu^a=\bar q_i\gamma_\mu(1-\gamma_5)\lambda^a
q_j$ has the chiral representation  (see e.g. \cite{Georgi})
 \be
 J_\mu^a= -{if_\pi^2\over 4}\,{\rm
 Tr}(U^\dagger\lambda^a\partial_\mu U-\partial_\mu
 U^\dagger\lambda^a U)=-{if_\pi^2\over 2}\,{\rm
 Tr}(U^\dagger\lambda^a\partial_\mu U).
 \en
It is straightforward to show that $J_\mu=\bar
q_i\gamma_\mu(1-\gamma_5)q_j$ has the expression
 \be
 J^{ji}_\mu=-{if_\pi^2\over 2}\left({2i\over f_\pi}\partial_\mu\phi+{2\over
 f_\pi^2}[\phi,\partial_\mu\phi]-{i\over
 f_\pi^3}a_3\{\phi^2,\partial_\mu\phi\}+{i\over
 f_\pi^3}(4-a_3)\phi\partial_\mu\phi\,\phi+\cdots\right)^{ji}.
 \en
Note that the sign convention of $J_\mu^a$ or $J_\mu$ is chosen in
such a way that $\la 0|J_\mu|P(p)\ra=-if_\pi p_\mu$. We are ready
to evaluate the point-like 3-body matrix element
 \be
 \la \pi^-(p_1)\pi^+(p_2)K^-(p_3)|(\bar su)_\vma|0\ra_{\rm contact}=-{i\over
 f_\pi}\left[ {a_3\over 2}(p_1+p_2+p_3)_\mu-2p_{2\mu}\right],
 \en
which is chiral-realization dependent. This realization dependence
should be compensated by the pole contribution, namely, the $B^-$
to $K^-$ weak transition followed by the strong interaction
$K^-\to K^-\pi^+\pi^-$. The strong vertex followed from the chiral
Lagrangian (\ref{chiLang}) has the form
 \be
 S=-{ia_3\over 2f_\pi^2}(p^2-m_3^2)+{2i\over f_\pi^2}\,p\cdot p_2,
 \en
with $p=p_1+p_2+p_3$. Hence,
 \be
 \la \pi^-(p_1)\pi^+(p_2)K^-(p_3)|(\bar su)_\vma|0\ra &=& \la
 \pi^-\pi^+K^-|(\bar su)_\vma|0\ra_{\rm
 contact}+S{i\over p^2-m_K^2}\la K^-(p)|(\bar su)_\vma|0\ra  \non \\
 &=& {2i\over f_\pi}\left(p_{2\mu}-{p\cdot p_2\over
 p^2-m_K^2}p_\mu\right).
 \en
Evidently, the $a_3$ terms are cancelled as it should be. It is
worth stressing again that the above matrix element is valid only
for low-momentum pseudoscalars. It is easily seen that in the
chiral limit
 \be \label{ann}
 \la \pi^-\pi^+K^-|(\bar su)_\vma|0\ra\la 0|(\bar u
 d)_\vma|B^-\ra=0.
 \en
Physically, the helicity suppression is perfect when light
final-state pseudoscalar mesons are massless. Although Eq.
(\ref{ann}) is derived for soft Goldstone bosons, it should hold
even for the energetic kaon and pions  as the helicity suppression
is expected to be more effective.

The factorizable contributions due to the penguin operator $O_6$
is
 \be \label{O6mea}
 \la \pi^-\pi^+K^-|O_6|B^-\ra &=& -2\Big\{\la K^-|\bar
 s(1+\gamma_5)u|0\ra\la\pi^-\pi^+|\bar u(1-\gamma_5)b|B^-\ra \non
 \\ &+& \la \pi^+K^-|\bar s(1+\gamma_5)d|0\ra\la\pi^-|\bar
 d(1-\gamma_5)b|B^-\ra \non \\
 &+& \la \pi^-\pi^+K^-|\bar s(1+\gamma_5)u|0\ra\la 0|\bar
 u(1-\gamma_5)b|B^-\ra\Big\}.
 \en
Applying equations of motion we obtain
 \be
 \la K^-|\bar
 s(1+\gamma_5)u|0\ra\la\pi^-\pi^+|\bar
 u(1-\gamma_5)b|B^-\ra &=& {m_K^2\over m_bm_s}\la K^-|(\bar s
 u)_\vpa|0\ra\la\pi^-\pi^+|(\bar ub)_\vpa|B^-\ra \non \\
 &=& {m_K^2\over m_bm_s}\la K^-|(\bar s
 u)_\vma|0\ra\la\pi^-\pi^+|(\bar ub)_\vma|B^-\ra,
 \en
and
 \be
 && \la \pi^+(p_2)K^-(p_3)|\bar s(1+\gamma_5)d|0\ra\la\pi^-(p_1)|\bar
 d(1-\gamma_5)b|B^-\ra \non \\
 && \qquad\qquad ={(p_2+p_3)^\mu\over m_s}\la
 \pi^+(p_2)K^-(p_3)|\bar s\gamma_\mu u|0\ra\,{m_B^2-m_\pi^2\over
 m_b}F_0^{B\pi}(t) \non \\
 && \qquad\qquad ={m_K^2-m_\pi^2\over m_s}\,{m_B^2-m_\pi^2\over
 m_b}\,F_0^{K\pi}(t)F_0^{B\pi}(t).
 \en
To evaluate the three-body matrix element $\la \pi^-\pi^+K^-|\bar
s(1+\gamma_5)u|0\ra$, we will first consider the case that the
kaon and pions are soft and then assign a form factor to account
for their momentum dependence. At low energies, it is known that
the light-to-light current can be expressed in terms of light
pseudoscalars (see e.g. \cite{Cheng88})
 \be
 \bar q_j(1-\gamma_5)q_i={f_\pi^2 v\over 2}U_{ij},
 \en
to the lowest order in the light meson derivatives, where
 \be
 v=\,{m_{\pi^+}^2\over m_u+m_d}=\,{m_{K^+}^2\over
 m_u+m_s}=\,{m_K^2-m_\pi^2\over m_s-m_d}
 \en
characterizes the quark-order parameter $\la\bar qq\ra$ which
spontaneously breaks the chiral symmetry. It is easily seen that
the point-like contact term yields
 \be
 \la \pi^-\pi^+K^-|\bar s\gamma_5 u|0\ra_{\rm
 contact}=\,i{a_3\over 2}{v\over f_\pi}.
 \en
As before, this chiral-realization dependence should be
compensated by the pole contribution, namely, the weak transition
of $B^-$ to $K^-$ followed by the strong scattering $K^-\to
K^-\pi^+\pi^-$. Hence,
 \be
 \la \pi^-(p_1)\pi^+(p_2)K^-(p_3)|\bar s\gamma_5 u|0\ra &=& \la
 \pi^-\pi^+K^-|\bar s\gamma_5 u|0\ra_{\rm
 contact}+S{i\over p^2-m_K^2}\la K^-(p)|\bar s\gamma_5 u|0\ra \non \\
 &=& {iv\over f_\pi}\left(1-{2p_1\cdot p_3\over
 m_B^2-m_K^2}\right).
 \en
Therefore, the $a_3$ terms are cancelled. Note that, contrary to
the $(V-A)(V-A)$ case where the weak annihilation vanishes in the
chiral limit, the penguin-induced weak annihilation does not
diminish in the same limit. This is so because the helicity
suppression works for the $(V-A)(V-A)$ interaction but not for the
$(S-P)(S+P)$ one.

Putting everything together leads to
 \be \label{O6me}
 \la \pi^-\pi^+K^-|O_6|B^-\ra &=& -2\Bigg\{ {m_K^2\over m_bm_s}\la K^-|(\bar s
 u)_\vma|0\ra\la\pi^-\pi^+|(\bar ub)_\vma|B^-\ra
 +{m_K^2-m_\pi^2\over m_s}\,{m_B^2-m_\pi^2\over
 m_b} \non \\ &\times&
 \Bigg[ F_0^{K\pi}(t)F_0^{B\pi}(t)-{f_Bf_K\over
 f_\pi^2}\left(1-{2p_1\cdot p_3\over
 m_B^2-m_K^2}\right)F^{K\pi\pi}(m_B^2)\Bigg]\Bigg\},
 \en
where  the form factor $F^{K\pi\pi}$ is needed to accommodate the
fact that the final-state pseudoscalars are energetic rather than
soft. The full amplitude finally reads
 \be \label{BKpipi}
 A(B^-\to \pi^-\pi^+K^-)_{\rm NR} &=& {G_F\over\sqrt{2}}\Bigg\{\Big(
 V_{ub}V_{us}^*a_1-V_{tb}V_{ts}^*\Big[
 a_4+a_{10}-2(a_6+a_8)\,{m_K^2\over m_bm_s}\Big]\Big) \non \\
 &\times& \la K^-|(\bar s u)_\vma|0\ra\la\pi^-\pi^+|(\bar
 ub)_\vma|B^-\ra \non \\
 &+& \Big[V_{ub}V_{us}^*a_2-V_{tb}V_{ts}^*{3\over 2}(a_7+
 a_9)\Big]\,F_1^{BK}(s)F^{\pi\pi}(s)(t-u) \non \\
 &-& V_{tb}V_{ts}^*\Bigg((a_4-{1\over 2}a_{10})\Big[F_0^{B\pi}(t)
 F_0^{K\pi}(t){(m_B^2-m_\pi^2)(m_K^2-m_\pi^2)\over t}  \non \\
 &+&  F_1^{B\pi}(t)F_1^{K\pi}(t)
(m_B^2+2m_\pi^2+m_K^2-2s-t-{(m_B^2-m_\pi^2)(m_K^2-m_\pi^2)\over
t}\Big] \non \\
&-& (2a_6-a_8){m_B^2-m_\pi^2\over m_b}\,{m_K^2-m_\pi^2\over
 m_s} \non \\
 &\times & \Big[ F_0^{B\pi}(t)F_0^{K\pi}(t)-{f_Bf_K\over
 f_\pi^2}\left(1-{2p_1\cdot p_3\over
 m_B^2-m_K^2}\right)F^{K\pi\pi}(m_B^2)\Big]\Bigg)\Bigg\},
 \en
where $u\equiv (p_B-p_2)^2$. As noted in passing, we should only
consider the pole contribution to the 3-body matrix element
$\la\pi^-\pi^+|(\bar ub)_\vma|B^-\ra$ so that
 \be \label{contact1}
 && \la K^-(p_3)|(\bar s u)_\vma|0\ra\la\pi^-(p_1)\pi^+(p_2)|(\bar
 ub)_\vma|B^-\ra_{\rm pole} \non \\
 &=& F_1^{B\pi}(m_K^2)\,{f_K\over f_\pi}\,{g\sqrt{m_Bm_{B^*}}\over t-m_{B^*}^2}
 \left[ m_B+{t\over m_B}-m_B{m_B^2-t\over
 m_K^2}\left(1-{F_0^{B\pi}(m_K^2)\over
 F_1^{B\pi}(m_K^2)}\right)\right]  \non \\
 &\times&
 \left[s+t-m_B^2-m_2^2+{(t-m_2^2+m_3^2)(m_B^2-t-m_1^2)\over
 2m_{B^*}^2}\right].
 \en

The decay amplitudes for other decays $B^-\to \pi^-(K^-)h^+h^-$
and $\ov B^0\to\ov K^0h^+h^-$ have the similar expressions as Eq.
(\ref{BKpipi}) except for $B^-\to \pi^+\pi^-\pi^-$ and $B^-\to
K^+K^-K^-$ where one also needs to add the contributions from the
interchange $s\leftrightarrow t$ and put a factor of 1/2 in the
decay rate to account for the identical particle effect.

\subsection{Results and discussions}
Before proceeding to the numerical results, it is useful to
express the direct 3-body decays of the heavy mesons in terms of
some quark-graph amplitudes \cite{CC90,CC87}: $\t_1$ and $\t_2$,
the color-allowed external $W$-emission tree diagrams; $\c_1$ and
$\c_2$, the color-suppressed internal $W$-emission diagrams; $\e$,
the $W$-exchange diagram; $\a$, the $W$-annihilation diagram;
${\cal P}_1$ and ${\cal P}_2$, the penguin diagrams, and ${\cal
P}_a$, the penguin-induced annihilation diagram. The quark-graph
amplitudes of various 3-body $B$ decays $B\to\pi h^+h^-$ and $B\to
Kh^+h^-$ are summarized in Table I. As mentioned in \cite{CC90},
the use of the quark-diagram amplitudes for three-body decays are
in general momentum dependent. This means that unless its momentum
dependence is known, the quark-diagram amplitudes of direct 3-body
decays cannot be extracted from experiment without making further
assumptions. Moreover, the momentum dependence of each
quark-diagram amplitude varies from channel to channel.

To consider the nonresonant contribution arising from the pion and
kaon electromagnetic form factors $F^{\pi\pi}$ and $F^{KK}$, we
follow \cite{Deshpande} to use the parametrization
 \be \label{em}
 F^{\rm em}_{\rm nonres}(q^2)=\,{1\over
 1-q^2/m_*^2+i\Gamma_*/m_*},
 \en
and employ $\Gamma_*=200$ MeV, and $m_*=600$ MeV for the pion and
700 MeV for the kaon. The momentum dependence of the weak form
factor $F^{K\pi}(q^2)$ is parametrized as
 \be \label{Kpi}
 F^{K\pi}(q^2)=\,{F^{K\pi}(0)\over 1-q^2/{\Lambda_\chi}^2+i\Gamma_* {\Lambda_\chi}},
 \en
where $\Lambda_\chi\approx 830$ MeV is the chiral-symmetry
breaking scale \cite{Cheng88}. Likewise, the form factor
$F^{K\pi\pi}$ appearing in Eq. (\ref{O6me}) is assumed to be
 \be \label{Kpipi}
 F^{K\pi\pi}(q^2)=\,{1\over 1-q^2/{\Lambda_\chi}^2}.
 \en

\begin{figure}[t]
\vspace{0cm} \hskip 3cm
  \psfig{figure=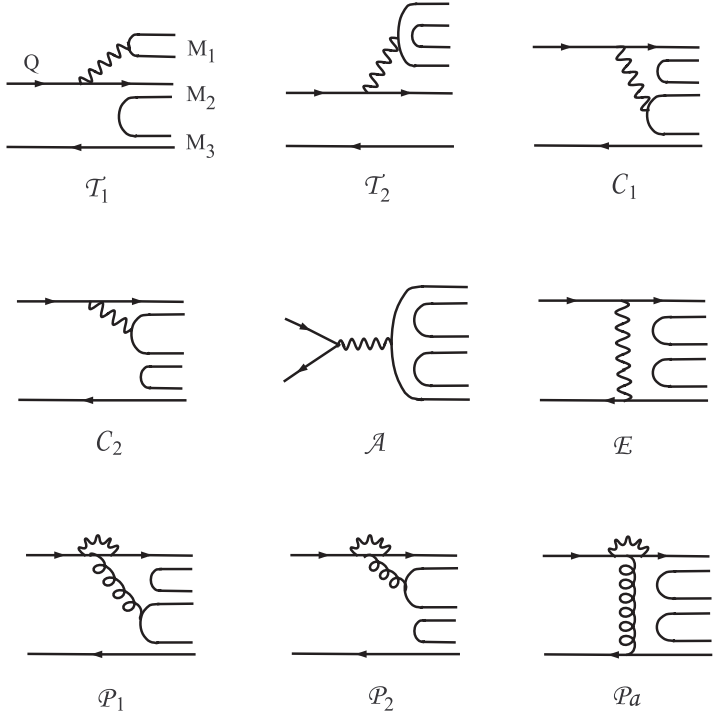,width=10cm}
\vspace{0.5cm}
    \caption[]{\small Quark diagrams for the three-body decays of heavy
    mesons, where $Q$ denotes a heavy quark.}
\end{figure}

The predicted branching ratios for direct charmless 3-body $B$
decays are shown in Table I. The decays $B^-\to \pi^-h^+h^-$ are
tree dominated and their main contributions come from the $B^*$
pole. In contrast, the decays $B^-\to (K^-h^+h^-)_{\rm NR}$ and
$\ov B^0\to(\ov K^0 h^+h^-)_{\rm NR}$ for $h=\pi,K$ are penguin
dominated. When $h=\pi$, the main contribution comes from the
2-body matrix elements of scalar densities, namely, the second
term on the right hand side of Eq. (\ref{O6mea}), while the
contribution from the three-body and one-body matrix elements of
pseudoscalar densities [the first term of Eq. (\ref{O6mea})]
characterized by the term $2a_6m_K^2/(m_bm_s)$ in Eq.
(\ref{BKpipi}) is largely compensated by the $a_4$ term.

Direct three-body charmless $B^\pm$ decays have been searched for
by CLEO \cite{CLEO} with limits summarized in Table~\ref{tab:B}.
The decays $B^-\to\pi^- K^+K^-,~K^-K^+K^-$ and $\ov B^0\to \ov
K^0\pi^+\pi^-,~\ov K^0K^+K^-$ were measured recently by Belle
\cite{Belle,Belle1} and BaBar \cite{BaBar} but without any
assumptions on the intermediate states. It is interesting to note
that the limits $1.2\times 10^{-5}$ set by Belle and $7\times
10^{-6}$ by BaBar for $\pi^-K^+K^-$ (resonant and nonresonant) is
improved over the previous CLEO limit $7.5\times 10^{-5}$ for the
nonresonant one. Needless to say, it is important to measure the
nonresonant decay rates by $B$ factories and compare them with
theory.

In the estimation of direct 3-body decay rates we have applied the
$B^*BP$ strong coupling given by Eq. (\ref{g}) and the $B^*\to P$
weak transition beyond their validity. Needless to say, this will
cause some major theoretical uncertainties in the calculations
because the strong $B^*BP$ coupling is derived under heavy quark
and chiral symmetries and hence the momentum of the soft
pseudoscalar should be less than $\Lambda_\chi$. For the energetic
pseudoscalar, the intermediate $B^*$ state is far from its mass
shell.
It is assumed in \cite{Deshpande} that the off-shellness of the
$B^*$ pole is accounted for by replacing the term $\sqrt{m_{B^*}}$
in Eq. (\ref{g}) by $(p_{B^*}^2)^{1/4}$ and it is found that the
branching ratios are reduced by $(30\sim 40)\%$ for $B^-\to \pi^-
K^+ K^-, \pi^+\pi^-\pi^-$ as shown in Table~\ref{tab:B}, while
$B^-\to K^-h^+h^-$ for $h=\pi,K$ remain essentially unaffected.
 Using the measured branching ratios $(55.6\pm
5.8\pm7.7)\times 10^{-6}$ and $(35.3\pm3.7\pm4.5)\times 10^{-6}$
by Belle \cite{Belle}, $(59.2\pm4.7\pm4.9)\times 10^{-6}$ and
$(34.7\pm2.0\pm 1.8)\times 10^{-6}$ by BaBar \cite{BaBar} for
$B^-\to K^-\pi^+\pi^-$ and $B^-\to K^-K^+K^-$, respectively, in
conjunction with  the calculated results for direct 3-body decays,
the corresponding fractions of nonresonant components are found to
be 4\% and 3\%, respectively.

\vskip 0.4cm {\squeezetable
\begin{table}[ht]
\caption{Quark-diagram amplitudes and branching ratios for
nonresonant 3-body charmless $B$ decays. The prediction $\B^1_{\rm
theor}$ is made for $g_{B B_{(s)}^* K(\pi)}=2g/f_\pi \times (m_B
m_{B_{(s)}^*})^{1/2}$ while the $\B^2_{\rm theor}$ accounts the
off-shellness of the $B_{(s)}^*$ by letting $g_{B B_{(s)}^*
K(\pi)}=2g/f_\pi \times \left(m_B
\sqrt{p_{B_{(s)}^*}^2}\right)^{1/2}$. Experimental limits are
taken from [31].}
\begin{center}
 \begin{tabular}{
l c l l l }
Decay mode & Quark-diagram amplitude & $\B^1_{\rm theor}$ & $\B^2_{\rm theor}$ & $\B_{\rm expt }$ \cite{PDG02} \\
\hline
 $B^-\to\pi^-\pi^+\pi^-$ & $V_{ub}V_{ud}^*\sqrt{2}(\t_1+\c_1+\a)+V_{tb}
 V_{td}^*\sqrt{2}(\p_1+\p_2+\p_a)$ & $3.0\times 10^{-6}$ & $1.7\times 10^{-6}$ & $<4.1\times 10^{-5}$ \\
 $\quad~\to \pi^-K^+K^-$ & $V_{ub}V_{ud}^*(\t_1+\c_1+\a)+V_{tb}V_{td}^*(\p_1+\p_2+\p_a)$ & $1.8\times
 10^{-6}$ & $1.3\times
 10^{-6}$ & $<7.5\times 10^{-5}$ \\
 $\quad~\to K^-\pi^+\pi^-$ & $V_{ub}V_{us}^*(\t_1+\c_1+\a)+V_{tb}V_{ts}^*(\p_1+\p_2+\p_a)$ &
 $2.4\times 10^{-6}$ &
 $2.3\times 10^{-6}$ & $<2.8\times 10^{-5}$  \\
 $\quad~\to K^-K^+K^-$ & $V_{ub}V_{us}^*\sqrt{2}(\t_1+\c_1+\a)+V_{tb}V_{ts}^*\sqrt{2}(\p_1+\p_2+\p_a)$ &
 $9.1\times 10^{-7}$ & $8.5\times 10^{-7}$ & $<3.8\times 10^{-5}$ \\
 $\ov B^0\to\ov K^0\pi^+\pi^-$ & $V_{ub}V_{us}^*\c_1+V_{tb}V_{ts}^*(\p_1+\p_2+\p_a)$
 & $2.1\times 10^{-6}$ & $2.1\times 10^{-6}$\\
 $\quad~\to\ov K^0K^+K^-$ & $V_{ub}V_{us}^*(\t_1+\c_1)+V_{tb}V_{ts}^*(\p_1+\p_2+\p_a)$
 & $1.2\times 10^{-6}$ & $1.2\times 10^{-6}$\\
\end{tabular}
\end{center}\label{tab:B}
\end{table}}

\section{Nonresonant three-body decays of $D$ mesons}
For nonresonant three-body $D$ decays, the applicability of HMChPT
should be in a  better position than the $B$ meson case. In Table
II the maximum momentum $p$ of any of the decay products in the
$D$ rest frame is listed. As stressed in \cite{Zhang}, $D\to KKK$
are the decay modes where HMChPT can be reliably applied since $p$
there is of order 545 MeV which is below the chiral symmetry
breaking scale. For other $\ov K\pi\pi$ and $\ov K K\pi$ modes,
the regime of the phase space where HMChPT is applicable is not
necessarily small.

The calculations for nonresonant three-body decays of the charmed
mesons proceed in the same way as the $B$ meson case and they are
performed in the framework of HMChPT for two different cases: (i)
HMChPT is applied to both strong and weak vertices, and (ii) it is
applied only to the strong vertex and the weak transition is
accounted for by  form factors. These two different cases are
denoted by $\B^a$ and $\B^b$, respectively, in Table II. Here we
would like to point out some interesting physics. First, consider
the decay $D^0\to \ov K^0\pi^+\pi^-$. In HMChPT its amplitude is
given by
 \be
 A(D^0\to \pi^-(p_1)\ov
 K^0(p_2)\pi^+(p_3))=-{G_F\over\sqrt{2}}V_{cs}V_{ud}^*(a_1A_1+a_2A_2),
 \en
with
 \be
 A_1 &=& {f_\pi\over 2}\left\{2m_3^2r+(m_D^2-s-m_3^2)\omega_++(2t+s-m_D^2-2m_2^2-m_3^2)
 \omega_-\right\},  \non \\
 A_2 &=& {f_K\over 2}\left\{2m_2^2r+(m_D^2-u-m_2^2)\omega_++(2t+u-m_D^2-2m_3^2-m_2^2)
 \omega_-\right\},
 \en
where the form factors $r,~\omega_+$ and $\omega_-$ have similar
expressions as Eq. (\ref{r&omega}). Since $a_1$ and $a_2$ in $D$
decays are opposite in signs [see Eq. (\ref{Da12})], it follows
that the decay rate is suppressed owing to the destructive
interference, see Table II.

However, when HMChPT is applied only to the strong vertex, the
main contribution to $D^0\to \ov K^0\pi^+\pi^-$ comes from the
$D^{*+}$ pole, namely, the strong process $D^0\to\pi^- D^{*+}$
followed by the weak transition $D^{*+}\to \ov K^0\pi^+$. Since it
is known that the interference in $D^+\to\ov K^0\pi^+$ is
destructive, naively it is expected that the same destructive
interference occurs in the nonresonant $D^0\to \ov K^0\pi^+\pi^-$
decay. However, this is not the case. The $D^*$ pole amplitude is
 \be
 A(D^0\to \ov K^0\pi^+\pi^-)_{\rm pole} &=&
 A^\mu_{D^*\pi K}\,{i(-g_{\mu\nu}+p_{D^*\mu}p_{D^*\nu}/m_{D^*}^2)\over
 p_{D^*}^2-m_{D^*}^2}\,A_{DD^*\pi}^\nu.
 \en
Now under factorization
 \be
 \vp_\mu A^\mu_{D^*\pi K} =
 {G_F\over\sqrt{2}}V_{cs}V_{ud}^*&\Big\{&a_1\la \pi^+(p_3)|(\bar
 ud)_\vma|0\ra\la\ov K^0(p_2)|(\bar sc)_\vma|D^{*+}(p_{D^*})\ra \non \\
 &+& a_2\la \ov K^0(p_2)|(\bar sd)_\vma|0\ra\la\pi^+(p_3)|(\bar
 uc)_\vma|D^{*+}(p_{D^*})\ra \Big\}.
 \en
Applying heavy quark symmetry one can relate the form factors in
$\la \ov K^0|(\bar sc)_\vma|D^{*+}\ra$ to those in $\la \ov
K^0|(\bar sc)_\vma|D^{+}\ra$:
 \be
 \la \ov K^0(p_K)|(\bar
 sc)_\vma|D^{+}(p_D)\ra=f_+^{DK}(q^2)(p_D+p_K)_\mu+f_-^{DK}(q^2)(p_D-p_K)_\mu.
 \en
We obtain
 \be
 \vp_\mu A^\mu_{D^*\pi K} = -i{G_F\over\sqrt{2}}V_{cs}V_{ud}^*(\vp\cdot
 p_3)&\Bigg\{& a_1\,f_\pi\Big[
 (f_++f_-)^{DK}m_D+(f_+-f_-)^{DK}{t\over m_D}\Big] \non \\
 &-& a_2\,f_K\Big[
 (f_++f_-)^{D\pi}m_D+(f_+-f_-)^{D\pi}{t\over m_D}\Big]\Bigg\}.
 \en
It is interesting to note that although the interference is
destructive in $D^{*+}\to\ov K^0\pi^+$, it becomes constructive in
the process $D^0\to\pi^-D^{*+}\to \pi^-\pi^+\ov K^0$. We see from
Table II that $\B^b$ is indeed much larger than $\B^a$ for $D^0\to
\ov K^0\pi^+\pi^-$.

\vskip 0.4cm
\begin{table}[th]
\caption{Quark-diagram amplitudes and branching ratios (in
percent) for nonresonant 3-body $D$ decays, where $p$ (in units of
MeV) is the largest momentum any of the products can have in the
$D$ rest frame. Heavy meson chiral perturbation theory is applied
to both heavy-light strong and weak vertices for the theoretical
prediction $\B^a$, while it is applied only to the strong vertex
for $\B^b$. Form factors for $D\to \pi$ and $D\to K$ transitions
are taken from [25] and experimental results  from [31]. For the
recent measurements of the nonresonant decays $D^+\to
K^-\pi^+\pi^+$, $D^0\to \bar K^0K^+K^-$ and
$D_s^+\to\pi^+\pi^+\pi^-$, see the text.}
\begin{center}
\begin{tabular}{  l l c l l l }
Decay mode & $p$ & Quark-diagram amplitude & $\B^a_{\rm theor}$ &
$\B^b_{\rm theor}$ & $\B_{\rm expt}$ \cite{PDG02}
\\ \hline  \\
 $D^0\to\ov K^0\pi^+\pi^-$ & 842 & $V_{ud}V_{cs}^*(\t_1+\c_2+\e)$ &
 0.03 & 0.17 & see~text    \\
 $\quad~\to K^-\pi^+\pi^0$ & 844 &
 $V_{ud}V_{cs}^*{1\over\sqrt{2}}(\t_1+\c_1)$ & 0.61 & 0.28 &$1.05^{+0.51}_{-0.19}$
  \\
  $\quad~\to \ov K^0K^+K^-$ & 544 & $V_{ud}V_{cs}^*(\t_2+\c_2+\e)$ & 0.16 &
  0.01 &  $0.55\pm0.09$ \\
  $D^+\to \ov K^0\pi^+\pi^0$ & 845 &
  $V_{ud}V_{cs}^*{1\over\sqrt{2}}(\t_1+\c_1)$ & 1.5 & 0.7 & $1.3\pm 1.1$ \\
  $\qquad\to K^-\pi^+\pi^+$ & 845 & $V_{ud}V_{cs}^*\sqrt{2}(\t_1+\c_1)$ & 6.5 & 1.6 &
  $8.6\pm0.8$ \\
  $\qquad\to\pi^+\pi^+\pi^-$ & 908 &
  $V_{ud}V_{cd}^*\sqrt{2}(\t_1+\c_1+\a+\p_1)+V_{us}V_{cs}^*\sqrt{2}(\p_1)$ & 0.50 & 0.067 &
  $0.024\pm 0.021$ \\
  $\qquad\to K^-K^+\pi^+$ & 744 &
  $V_{ud}V_{cd}^*(\a+\p_1)+V_{us}V_{cs}^*(\t_1+\c_1+\e)$ & 0.48 & 0.004 &
  $0.45\pm 0.09$ \\
  $D_s^+\to K^-K^+\pi^+$ & 805 & $V_{ud}V_{cs}^*(\t_1+\c_1+\a)$ & 1.0 & 0.69 &
  $0.9\pm 0.4$ \\
  $\qquad\to\pi^+\pi^+\pi^-$ & 959 & $V_{ud}V_{cs}^*\sqrt{2}(\a)$ & & &
  $0.005\pm0.022$ \\
\end{tabular}
\end{center}
\end{table}

The nonresonant decay $D^0\to (\ov K^0K^+K^-)_{\rm NR}$ deserves a
special attention for two reasons. First, it is the only
Cabibbo-allowed direct 3-body mode which receives contributions
from the external $W$-emission diagram $\t_2$ (see Fig. 2).
Second, as noted in passing, HMChPT is presumably most reliable
for this mode.  Its factorizable amplitude has the form
 \be
 A(D^0\to K^-(p_1)K^+(p_2)\ov K^0(p_3))_{\rm NR} &=&
 {G_F\over\sqrt{2}}V_{ud}V_{cs}^*\Bigg\{
 a_1\la K^+\ov K^0|(\bar ud)_\vma|0\ra\la K^-|(\bar
 sc)_\vma|D^0\ra  \non \\
 &+& a_2\la \ov K^0|(\bar sd)_\vma |0\ra\la K^-K^+|(\bar
 uc)_\vma|D^0\ra \non \\
 &+& a_2\la K^-K^+\ov K^0|(\bar sd)_\vma|0\ra\la 0|(\bar
 uc)_\vma|D^0\ra
 \Bigg\},
 \en
where the three terms on the right hand side correspond to the
quark diagrams $\t_2$, $\c_2$ and $\e$, respectively. Proceeding
as before and neglecting the $W$-exchange contribution in the
chiral limit, we obtain
  \be
 && A(D^0\to K^-(p_1)K^+(p_2)\ov K^0(p_3))_{\rm NR}
 = {G_F\over\sqrt{2}}V_{ud}V_{cs}^*\Big\{a_1
 A'_1+a_2A'_2\Big\},
 \en
where
 \be \label{A2}
 A'_1 &=& {f_{D}\over f_\pi}\left\{
 {g\sqrt{m_Dm_{D_s^*}^3}\over t-m_{D_s^*}^2}-{1\over 2}\right\}(s-u), \non \\
 A'_2 &=& -{f_K\over 2} \left\{2m_3^2r+(m_D^2-s-m_3^2)\omega_++(2t+s-m_D^2-2m_2^2-m_3^2)
 \omega_-\right\},
 \en
when HMChPT is applied to both strong and weak vertices, or
 \be
 A'_1 &=& F_1^{DK}(t)F^{KK}(t)(s-u), \non \\
 A'_2&=& F_1^{D_sK}(m_K^2){g\sqrt{m_Dm_{D_s^*}}\over t-m_{D_s^*}^2}
 \left[ m_D+{t\over m_D}-m_D{m_D^2-t\over
 m_K^2}\left(1-{F_0^{D_sK}(m_K^2)\over
 F_1^{D_sK}(m_K^2)}\right)\right]  \non \\
 &\times&
 \left[s+t-m_D^2-m_2^2+{(t-m_2^2+m_3^2)(m_D^2-t-m_1^2)\over
 2m_{D_s^*}^2}\right],
 \en
when HMChPT is applied only to the strong vertex. Again, the form
factors $r$, $\omega_+$ and $\omega_-$ in Eq. (\ref{A2}) have the
similar expressions as Eq. (\ref{r&omega}).

It is clear from Table II that the predicted branching ratio
$\B^a$ of 0.16\% for $D^0\to(\ov K^0K^+K^-)_{\rm NR}$ works much
better than $\B^b$, though the former is still too small compared
to the experimental value $(0.55\pm 0.09)\%$ \cite{PDG02}.  This
decay was also considered by Zhang \cite{Zhang} within the same
framework of HMChPT, but his result $2.3\times 10^{-4}$ for the
branching ratio, which is similar to the prediction $2\times
10^{-4}$ based on chiral perturbation theory \cite{CC90}, is
smaller than ours by one order of magnitude.

Some simple relations among different modes follow from the quark
diagram approach. For example, neglecting the weak annihilation
and penguin contributions and the phase space difference among
different modes, it is expected that
 \be
 {\B(D^+\to\pi^+\pi^+\pi^-)_{\rm NR}\over \B(D^+\to \pi^+\pi^+K^-)_{\rm NR}} &=&
 \left|{V_{cd}\over V_{cs}}\right|^2,   \non \\
 {\B(D^+\to \ov K^0\pi^+\pi^0)_{\rm NR}\over \B(D^+\to K^-\pi^+\pi^-)_{\rm NR}} &=&
 {1\over 4},   \\
 {\B(D^+\to K^-K^+\pi^+)_{\rm NR}\over \B(D^+\to \pi^+\pi^+\pi^-)_{\rm NR}} &=&
 {1\over 2},  \non \\
 {\B(D_s^+\to K^-K^+\pi^+)_{\rm NR}\over \B(D^+\to K^-\pi^+\pi^+)_{\rm NR}} &=&
 {1\over 2}{\tau(D_s^+)\over \tau(D^+)}. \non
 \en
The above anticipation can be checked against the experimental
results. It is easily seen that the measured $D^+\to
(\pi\pi\pi)_{\rm NR}$ is too small compared to the theoretical
prediction. For example, the observation that $(\pi^+K^+K^-)_{\rm
NR}\gg (\pi^+\pi^+\pi^-)_{\rm NR}$ in $D^+$ decays is rather
unexpected.

We see from Table II that the predictions for case (i) denoted by
$\B^a$ are generally larger than case (ii) denoted by $\B^b$
except for the decay $D^0\to\ov K^0\pi^+\pi^-$. Contrary to the
$B$ meson case where the predicted rates in these two different
methods can differ by one to two orders of magnitude, $\B^a$ and
$\B^b$ in some of $D$ decays differ only by a factor of 2. It is
also evident that in general $\B^b$s give a better agreement with
experiment for many of the direct 3-body $D$ decays , whereas
$\B^a$ works better for $D^0\to \ov K^0K^+K^-$ and $D^+\to
K^-K^+\pi^+$, though the prediction of the former mode by HMChPT
is still too small compared to experiment. As noted in the
Introduction, the early predictions based on SU(4) chiral
perturbation theory are in general too small when compared with
experiment \cite{CC90,Botella}.

There exist several new measurements of direct 3-body $D$ decays
in past few years: $D^0\to K^-\pi^+\pi^0,~\ov K^0\pi^+\pi^-,~\ov
K^0K^+K^-$, $D^+\to \pi^+\pi^+\pi^-,~K^-\pi^+\pi^+$ and
$D_s^+\to\pi^+\pi^+\pi^-$. The nonresonant branching ratio for the
first mode is found to be $(1.0\pm0.1\pm0.1^{+0.8}_{-0.1})\times
10^{-2}$ by CLEO \cite{CLEO1}. Previous experiments \cite{E687a}
indicate that the decay $D^+\to K^-\pi^+\pi^+$ is strongly
dominated by the nonresonant term with $(95\pm 7)\%$ \cite{PDG02}.
However, a recent Dalitz plot analysis by E791 \cite{E791a}
reveals that a best fit to the data is obtained if the presence of
an additional scalar resonance $\kappa$ is included. As a
consequence, the nonresonant decay fraction drops from 95\% to
$(13\pm 6)\%$, whereas $\kappa\pi^+$ accounts for $(48\pm12)\%$ of
the total rate. Therefore, the branching ratio of the direct decay
$D^+\to K^-\pi^+\pi^+$ is dropped from $(8.6\pm0.8)\%$ to
$(1.2\pm0.6)\%$. Likewise, It was found by the E687 experiment
that the decay $D^+\to \pi^+\pi^+\pi^-$ is dominated by the
nonresonant contribution with $(60\pm11)\%$ \cite{E687b}. Again,
the new Dalitz plot analysis by E791 \cite{E791b} points out that
half of the decays is accounted for by the scalar resonance
$\sigma$, whereas the nonresonant fraction is only $(7.8\pm
6.0\pm2.7)\%$. Consequently, $\B(D^+\to\pi^+\pi^+\pi^-)_{\rm NR}$
drops to $(0.024\pm 0.021)\%$. Very recently BaBar has reported
the preliminary result of the Dalitz plot analysis of $D^0\to\ov
K^0K^+K^-$ \cite{BaBarDKKK}. Its nonresonant fraction is estimated
to be $(0.4\pm0.3\pm0.8)\%$ and hence is negligible.

As for the direct decay $D^0\to \ov K^0\pi^+\pi^-$, the 2000
edition of Particle Data Group (PDG) \cite{PDG00} quotes a value
of $(1.47\pm 0.24)\%$ for its branching ratio. However, it is no
longer cited in the 2002 PDG \cite{PDG02} as no evidence for a
nonresonant component is seen according to the most detailed
analyses performed in \cite{Frabetti}. This is also confirmed by a
very recent CLEO measurement of this decay mode which gives
$(0.9\pm0.4^{+1.0+1.7}_{-0.3-0.2})\%$ for the nonresonant fraction
\cite{CLEODKpipi}.

The Cabibbo-suppressed decay $D_s^+\to(\pi^+\pi^+\pi^-)_{\rm NR}$
proceeds only through the $W$-annihilation diagram.  The early
E691 measurement gives $R=\B(D_s^+\to\pi^+\pi^+\pi^-)_{\rm
NR}/\B(D_s^+\to\phi\pi^+)=0.29\pm 0.09\pm0.03$ \cite{E691}.
However, it was found to be negligible by E791 \cite{E791c} and
its branching ratio is quoted to be $(5\pm 22)\times 10^{-5}$ by
2002 PDG (see Table II). Recently, FOCUS has reported the
preliminary result: the nonresonant fraction is measured to be
$(25.5\pm 4.6)\%$ \cite{FOCUS}. This corresponds to
$\B(D_s^+\to\pi^+\pi^+\pi^-)_{\rm NR}=(2.6\pm 0.9)\times 10^{-3}$.
Although the short-distance $W$-annihilation vanishes in the
chiral limit, the long-distance one can be induced from
final-state rescattering (see e.g. \cite{Cheng02}). \footnote{For
previous theoretical estimates, see \cite{Ivanov} and
\cite{Hoang}.} Therefore, the observation of direct
$D_s^+\to\pi^+\pi^+\pi^-$ implies the importance of final-state
interactions for nonresonant decays.

\section{Conclusions}
We have presented a systematical study of nonresonant three-body
decays of $D$ and $B$ mesons. We first draw some conclusions from
our analysis and then proceed to discuss the sources of
theoretical uncertainties during the course of calculation.

\begin{enumerate}
\item It is pointed out that if heavy meson chiral perturbation
theory (HMChPT) is applied to the heavy-light strong and weak
vertices and assumed to be valid over the whole kinematic region,
then the predicted decay rates for nonresonant 3-body $B$ decays
will be too large and especially $B^-\to \pi^-K^+K^-$ exceeds
substantially the current experimental limit. This can be
understood because chiral symmetry has been applied twice beyond
its region of validity.
 \item If HMChPT is applied only to the
strong vertex and the weak transition is accounted for by the form
factors, the dominant $B^*$ pole contribution to the
tree-dominated direct three-body $B$ decays will become small and
the branching ratio will be of order $10^{-6}$. The decay modes
$B^-\to (K^-h^+h^-)_{\rm NR}$ and $\ov B^0\to(\ov K^0 h^+h^-)_{\rm
NR}$ for $h=\pi,K$ are penguin dominated.
 \item We have considered
the use of HMChPT in two different cases to study the direct
3-body $D$ decays. We found that when HMChPT is applied only to
the strong vertex, the predictions in general give a better
agreement with experiment except for the decays $D^0\to K^- \pi^+
\pi^0, \ov K^0 K^+ K^-$ and $D^+\to K^-K^+\pi^+$ where a full use
of HMChPT to the weak vertices gives a better description. The
$D^{*+}$ pole contribution to $D^0\to\ov K^0\pi^+\pi^-$ proceeds
through external and internal $W$-emission diagrams with
constructive interference. The experimental observation that
$(\pi^+K^+K^-)_{\rm NR}\gg (\pi^+\pi^+\pi^-)_{\rm NR}$ in $D^+$
decays is largely unanticipated.

\end{enumerate}

It is useful to summarize the theoretical uncertainties
encountered in the present paper, though most of them have been
discussed before:
 \begin{itemize}
 \item For $B^*$ (and also $D^*$) pole contributions, the
 intermediate state $B^*$ is off its mass shell when the
 pseudoscalar meson coupled to $B^*$ and $B$ is no longer soft. This will affect
 the $B^*BP$ strong coupling.
 To estimate the off-shell effect of $B^*$, we replace its mass
 $m_{B^*}$ by $\sqrt{p_{B^*}^2}$ and find that the branching ratios
 for $B^-\to \pi^- K^+ K^-, \pi^+\pi^-\pi^-$ are reduced by $(
 30\sim 40)\%$, while $B^-\to K^-\pi^+\pi^-, K^-K^+K^-$ remain
 essentially unaffected.

 \item  We have parametrized
the $q^2$ dependence of the form factors $F^{\pi\pi}_{\rm nonres},
F^{KK}_{\rm nonres}$, $F^{K\pi}_{\rm nonres}$ and $F^{K\pi\pi}$ in
the form of Eqs. (\ref{em}), (\ref{Kpi}) and (\ref{Kpipi}).
However, part of scalar resonance effects is included in the
parametrization of the form factors. In the $B$ decays, the major
uncertainty of the calculated amplitudes comes from the chiral
enhanced term $\sim F_{0,{\rm nonres}}^{K\pi} (2a_6-a_8)\times
m_B(m_K^2-m_\pi^2)/m_s$. We may overestimate the penguin-dominant
nonresonant branching ratios if there exist scalar resonances,
e.g. $\kappa$. Although in some channels the $\sigma$ resonance is
included in $F^{\pi\pi}_{\rm nonres}$, its effect is suppressed by
the Cabibbo angle and by the fact that it decouples to the vector
current in the SU(2) symmetry limit.

 \item The point-like contact contribution to the three-body matrix element
 beyond the chiral limit,
e.g. $\la P_1P_2|(\bar qb)_\vma|B\ra_{\rm contact}$,  is unknown
but it becomes even smaller when $P_1$ or $P_2$ is not soft owing
to the smaller wave function overlap among $P_1, P_2$ and $B$.
Therefore it can be neglected in our calculations.

 \item Thus far we have assumed the factorization approximation to
 evaluate the decay amplitudes. It is known in the QCD
 factorization approach \cite{BBNS} that factorization is
 justified in the heavy quark limit where power corrections of
 order $1/m_B$ and $1/m_D$ can be neglected. Beyond the heavy
 quark limit, factorization is violated by power corrections which in general cannot
 be systematically explored. Nevertheless, some of them
 are calculable.
 For example, in the $B$ decays we have included the terms proportional to
 $a_6$ and $a_8$ which are of order $\bar\Lambda/m_b$ but chirally enhanced.
 Final-state interactions which have been neglected so far are also of order $\bar\Lambda/m_Q$.
 The decay $D_s^+\to(\pi^+\pi^+\pi^-)_{\rm NR}$ proceeds only through
 the $W$-annihilation process. Even if the short-distance contribution
 to the weak annihilation vanishes, it may receive sizable long-distance
 contributions via final-state rescattering. The preliminary FOCUS
 measurement of this mode may provide the first indication of the importance of final-state interactions
 for the weak annihilation process in nonresonant $D$ decays.
 A precise measurement of this mode can test the validity of the
 applying the factorization picture to the nonresonant three-body
 decays.
 \end{itemize}

\vskip 2.0cm \acknowledgments  H.Y.C. wishes to thank the C.N.
Yang Institute for Theoretical Physics at SUNY Stony Brook for its
hospitality. K.C.Y. would like to thank the Theory Group at the
Institute of Physics at Academia Sinica, Taipei for its
hospitality. This work was supported in part by the National
Science Council of R.O.C. under Grant Nos. NSC90-2112-M-001-047
and NSC90-2112-M-033-004.



\newpage
\begin{thebibliography}{99}
\newcommand{\bi}{\bibitem}

\bi{Deshpande} N.G. Deshpande, G. Eilam, X.G. He, and J.
Trampeti\'c, \pr {\bf D52}, 5354 (1995).

\bi{Fajfer1} S. Fajfer, R.J. Oakes, and T.N. Pham, \pr {\bf D60},
054029 (1999).

\bi{Fajfer2} B. Bajc, S. Fajfer, R.J. Oakes, T.N. Pham, and S.
Prelovsek, \pl {\bf B447}, 313 (1999).

\bi{Deandrea1} A. Deandrea, R. Gatto, M. Ladisa, G. Nardulli, and
P. Santorelli, \pr {\bf D62}, 036001 (2000); {\it ibid.} {\bf 62},
114011 (2000).

\bi{Deandrea} A. Deandrea and A.D. Polosa, \prl {\bf 86}, 216
(2001).

\bi{Gardner} J. Tandean and S. Gardner, hep-ph/0204147.

\bi{Bediaga} I. Bediaga, C. G\"obel, and R. M\'endez-Galain, \pr
{\bf D56}, 4268 (1997).

\bi{Singer} P. Singer, \pr {\bf D16}, 2304 (1977); {\sl Nuovo
Cim.} {\bf 42A}, 25 (1977).

\bi{KP} Yu. L. Kalinovsk and V.N. Pervushin, {\sl Sov. J. Nucl.
Phys.} {\bf 29}, 225 (1979).

\bi{Cheng86} H.Y. Cheng, \zp {\bf C32}, 243 (1986).

\bi{CC90} L.L. Chau and H.Y. Cheng, \pr  {\bf D41}, 1510 (1990).

\bi{Botella} F.J. Botella, S. Noguera, and J. Portol\'es, \pl {\bf
B360}, 101 (1995).

\bi{Yan} T.M. Yan, H.Y. Cheng, C.Y. Cheung, G.L. Lin, Y.C. Lin,
and H.L. Yu, \pr {\bf D46}, 1148 (1992); {\bf 55}, 5851(E) (1997).

\bi{Wise} M.B. Wise, \pr {\bf D45}, 2118 (1992).

\bi{Burdman} G. Burdman and J.F. Donoghue, \pl {\bf B280}, 287
(1992).

\bi{Zhang} D.X. Zhang, \pl {\bf B382}, 421 (1996).

\bi{Ivanov} A.N. Ivanov and N.I. Troitskaya, {\sl Nuovo Cim.} {\bf
A111}, 85 (1998).

\bi{Belle} Belle Collaboration, A. Garmash {\it et al.,} \pr {\bf
D65}, 092005 (2002).

\bi{BaBar} BaBar Collaboration, B. Aubert {\it et al.,}
hep-ex/0206004.

\bi{Du} D.S. Du, H. Gong, J. Sun, D. Yang, and G. Zhu, \pr {\bf
D65}, 094025 (2002).



\bi{Fajfer3} S. Fajfer, R.J. Oakes, and T.N. Pham, \pl {\bf B539},
67 (2002).

\bi{Lee} C.L.Y. Lee, M. Lu, and M.B. Wise, \pr {\bf D46}, 5040
(1992).

\bi{Cheng88} H.Y. Cheng, {\sl Int. J. Mod. Phys.} {\bf A4}, 495
(1989).

\bi{CLEOg} CLEO Collaboration, A. Anastassov {\it et al.,} \pr
{\bf D65}, 032003 (2002).

\bi{CLEO} CLEO Collaboration, T. Bergfeld {\it et al.,} \prl {\bf
77}, 4503 (1996).

\bibitem{BSW} M. Wirbel, B. Stech, and M. Bauer, \zp {\bf C29}, 637 (1985);
M. Bauer, B. Stech, and M. Wirbel, {\sl ibid.}  {\bf C34}, 103
(1987).

\bi{MS} D. Melikhov and B. Stech, \pr {\bf D62}, 014006 (2001).

\bi{Georgi} H. Georgi, {\it Weak Interactions and Modern Particle
Theory} (Benjamin/Cummings, New York, 1984).

\bi{CC87} L.L. Chau and H.Y. Cheng, \pr {\bf D36}, 137 (1987); \pl
{\bf B222}, 285 (1989).

\bi{Belle1} Belle Collaboration, talk presented by H.C. Huang at
XXXVIIth Rencontres de Moriond Electroweak Interactions and
Unified Theories (9 - 16 March 2002, Les Arcs, France).

\bi{PDG02} Particle Data Group,  K. Hagiwara {\it et al.,} \pr
{\bf D66}, 010001 (2002).

\bi{CLEO1} CLEO Collaboration, S. Kopp {\it et al.,} \pr {\bf
D63}, 092001 (2001).

\bi{E687a} E687 Collaboration, P.L. Frabetti {\it et al.,} \pl
{\bf B331}, 217 (1994); E691 Collaboration, J.C. Anjos {\it et
al.,} \pr {\bf D48}, 56 (1993); Mark III Collaboration, J. Adler
{\it et al.,} \pl {\bf B191}, 318 (1987).

\bi{E791a} E791 Collaboration, E.M. Aitala {\it et al.,}
hep-ex/0204018.

\bi{E687b} E687 Collaboration, P.L. Frabetti {\it et al.,} \pl
{\bf 407}, 79 (1997).

\bi{E791b} E791 Collaboration, E.M. Aitala {\it et al.,} \prl {\bf
86}, 770 (2001).

\bi{BaBarDKKK} BaBar Collaboration, B. Aubert {\it et al.,}
hep-ex/0207089.

\bi{PDG00} Particle Data Group, D.E. Groom {\it et al.,} {\sl Eur.
Phys. J.} {\bf C15}, 1 (2000).

\bi{Frabetti} P.L. Frabetti {\it et al.,} \pl {\bf B331}, 217
(1994); H. Albrecht {\it et al.,} {\sl ibid.} {\bf B308}, 435
(1993).

\bi{CLEODKpipi} CLEO Collaboration, H. Muramatsu {\it et al.,}
hep-ex/0207067.

\bi{E691} E691 Collaboration, J.C. Anjos {\it et al.,} \prl {\bf
62}, 125 (1989).

\bi{E791c} E791 Collaboration, E.M. Aitala {\it et al.,} \prl {\bf
86}, 765 (2001).

\bi{FOCUS} FOCUS Collaboration, talk presented by S. Erba at the
DPF Meeting of the American Physical Society at Williamsburgh,
Virginia, May 24-28, 2002.

\bi{Hoang} N.L. Hoang, A.V. Nguyen, and X.Y. Pham, \pl {\bf B357},
177 (1995).

\bi{Cheng02} H.Y. Cheng, hep-ph/0202254.

\bi{BBNS} M. Beneke, G. Buchalla, M. Neubert, and C.T. Sachrajda,
\np {\bf B591}, 313 (2000).


\end{thebibliography}
\end{document}